\documentclass[times,twocolumn]{aastex631}

\usepackage{graphicx}

\usepackage{enumitem}
\usepackage{amsmath}
\usepackage{multirow}
\usepackage{booktabs}
\usepackage{array}
\usepackage{bm}
\usepackage{textgreek}
\usepackage{etoolbox}

\usepackage{CJK}

\received{September 27, 2021}
\revised{October 18, 2021}
\accepted{October 20, 2021}

\shorttitle{Bald Patch Evolution}
\shortauthors{Lee et al.}

\begin{document}

\begin{CJK*}{UTF8}{gbsn}

\title{Rapid Evolution of Bald Patches in a Major Solar Eruption}

\author[0000-0001-7626-8386]{Jonathan H. Lee}
\affil{Institute for Astronomy, University of Hawai`i at M\={a}noa, 2680 Woodlawn Drive, Honolulu, HI 96822, USA}

\author[0000-0003-4043-616X]{Xudong Sun (孙旭东)}
\affil{Institute for Astronomy, University of Hawai`i at M\={a}noa, 34 Ohia Ku St, Pukalani, HI 96768, USA; \href{mailto:xudongs@hawaii.edu}{xudongs@hawaii.edu}}

\author[0000-0001-8975-7605]{Maria D. Kazachenko}
\affil{Department of Astrophysical and Planetary Sciences, University of Colorado Boulder, 2000 Colorado Ave, Boulder, CO 80305, USA}
\affil{National Solar Observatory, University of Colorado Boulder, 3665 Discovery Drive, Boulder, CO 80303, USA}



\begin{abstract}
Bald patch (BP) is a magnetic topological feature where U-shaped field lines turn tangent to the photosphere. Field lines threading the BP trace a separatrix surface where reconnection preferentially occurs. Here we study the evolution of multiple, strong-field BPs in active region 12673 during the most intense, X9.3 flare of solar cycle 24. The central BP, located between the initial flare ribbons, largely ``disintegrated'' within $35$ minutes. The more remote, southern BP survived. The disintegration manifested as a $9\degr$ rotation of the median shear angle; the perpendicular component of the horizontal field (with respect to the polarity inversion line) changed sign. The parallel component exhibited a step-wise, permanent increase of $1$~kG, consistent with previous observations of the flare-related ``magnetic imprint''. The observations suggest that magnetic reconnection during a major eruption may involve entire BP separatrices, leading to a change of magnetic topology from BPs to sheared arcades.
\end{abstract}




\section{Introduction}
\label{sec:intro}

The coronal magnetic fields drive explosive solar activities such as flares and coronal mass ejections. As excess magnetic energy accumulates, a magnetic flux rope may form, where field lines coherently wrap around a common axis for more than one turn. The flux rope is thought to be the central engine of solar eruptions \citep{forbes2000}.

In the core of solar active regions (ARs), the photospheric magnetic vectors, $\bm{B}$, typically point from the positive to the negative side across the polarity inversion line (PIL). The low-lying field lines form $\Omega$-shaped, sheared arcades that obliquely straddle the PIL. In some cases, however, they may also point from the negative to the positive side. The field lines are expected to be U-shaped and tangentially touch the photosphere at the PIL. These segments of the PIL are known as bald patches \citep[BPs;][]{titov1993}. BPs can occur in a variety of magnetic conditions, including quadrupolar configurations or potential fields. Shear BPs naturally occur when a magnetic flux rope lies low in the corona with its edge touching the photosphere.

Field lines threading the BP outline a separatrix surface, which is one of the magnetic topological features in the solar corona \citep{bungey1996}. If the BP field lines are ``tied'' at the photospheric PIL, current sheets may form at the separatrix as a response to the boundary motions \citep{low1988,billinghurst1993}. In reality, these field lines do not penetrate deeply into the photosphere. If the line-tying is relaxed, the BP separatrix effectively transforms into a quasi-separatrix layer \citep{demoulin1996}. A current layer of finite width may form, and the U-shaped field lines can gradually lift up \citep{karpen1990,karpen1991}.

Numerical simulations show that BPs can form during the emergence of a magnetic flux rope \citep{fan2004,archontis2009tri} or during flux cancellation \citep{vanballegooijen1989,aulanier2010}. Observationally, BPs can be identified in photospheric vector magnetograms. They are often found in emerging ARs \citep{okamoto2008,lites2010,kuckein2012}, decaying ARs \citep{yardley2016}, or quiet-Sun filament channels \citep{lopezariste2006}. The morphology of the X-ray sigmoids supports the existence of BPs in ARs with significant flux cancellation \citep{green2007,mckenzie2008}.


\begin{figure*}[t!]
\centering
\includegraphics[width=0.45\textwidth]{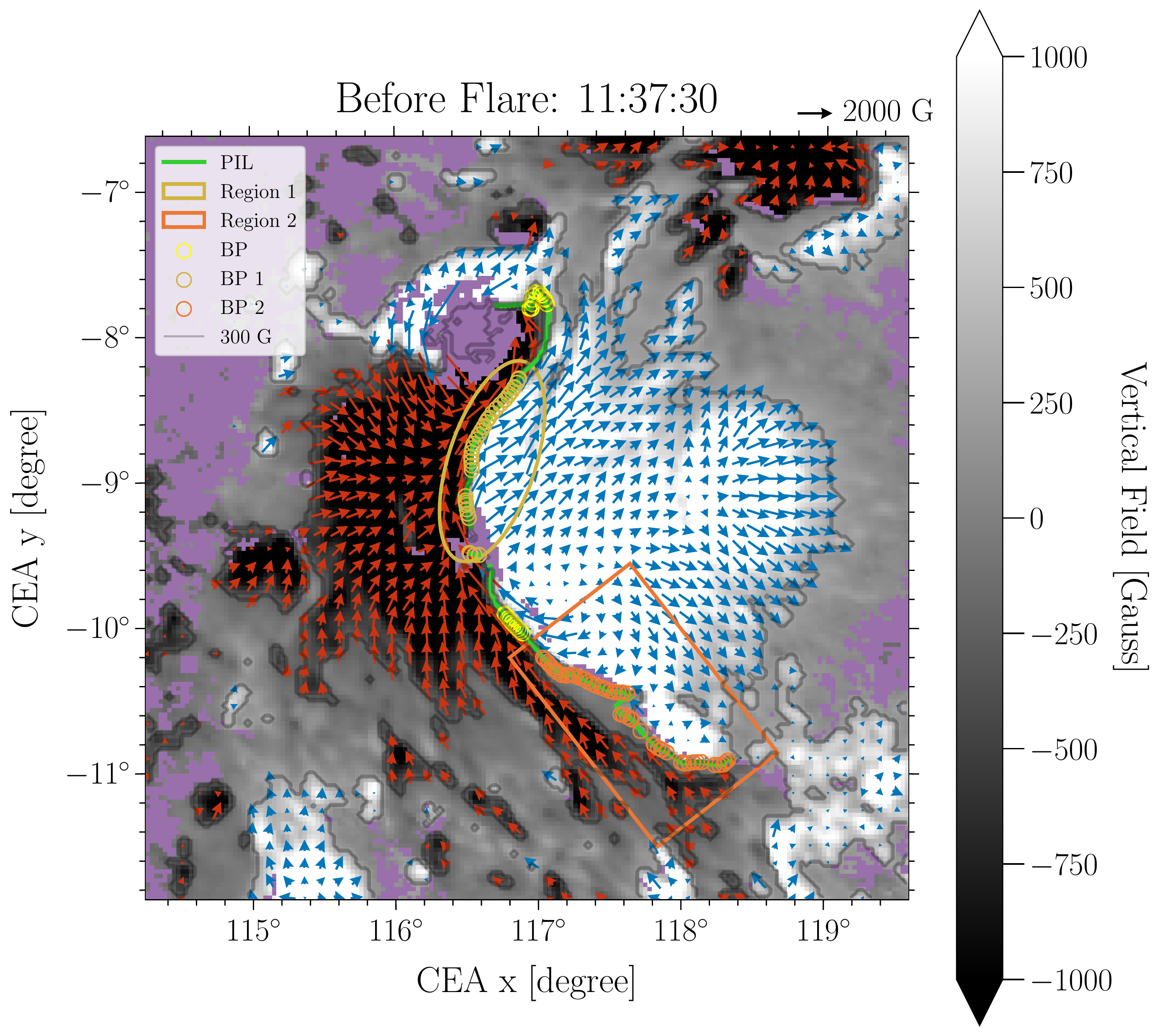}
\hspace{8mm}
\includegraphics[width=0.45\textwidth]{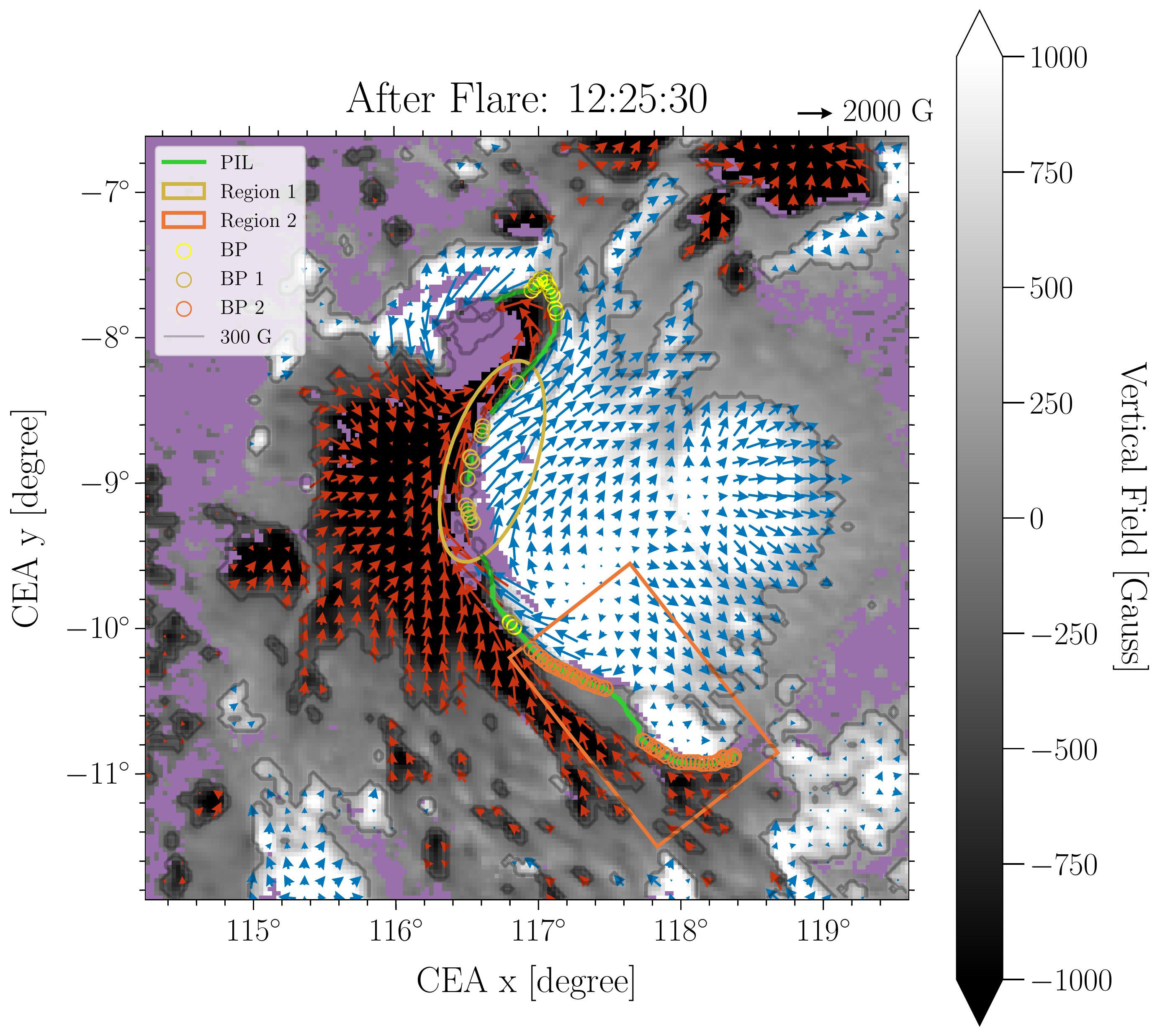}
\caption{AR 12673 before (11:37:30 UT) and after (12:25:30 UT) the flare. The background shows the positive (negative) polarity $B_{z}$ in white (black). Vectors show $\bm{B}_h$ with foot points in the positive (negative) $B_z$ in blue (red). The PIL and BPs are shown by the green line and yellow/orange circles, respectively. The gray contour is for $B_{z} = 300$~G. Purple color shows pixels that are excluded from our analysis due to weaker fields or larger uncertainties. The ellipse and the box indicate our selected regions of BP evolution analysis: Region 1 for BP disintegration, and Region 2 for BP survival. (An animation of this figure is available via \href{https://doi.org/10.5281/zenodo.5585623}{Zenodo}.)}
\label{fig:bpevo}
\end{figure*}


Coronal field models are often used to probe the magnetic topology above BPs. Using vector magnetograms as input, these models typically recover low-lying flux ropes, just as expected \citep{canou2009,guoy2010,yelleschaouche2012,chengx2014,liur2014,jiang2017,liulj2019}. BPs have also been inferred from coronal field models that employ line-of-sight magnetograms \citep{delannee1999,fletcher2001,wangtj2002,aulanier2002,mandrini2002}.

The photospheric magnetic fields may show rapid and permanent changes during flares \citep{sudol2005,wanghm2010,petrie2012,castellanosduran2018}. These ``magnetic imprints'' are now frequently observed with routine, full-disk vector magnetograms such as those from the Helioseismic and Magnetic Imager \citep[HMI;][]{schou2012,hoeksema2014}. The high-cadence ($90$ or $135$~s) HMI data, in particular, have revealed a structured evolution pattern \citep{sun2017a}. In an archetypical event, the horizontal field strength $B_h$ near the main PIL increased significantly during the earlier phase of the flare with a timescale of several minutes, while $B_h$ in the periphery decreased at later times with smaller magnitudes and a longer timescale. A potential explanation entails Lorentz force feedback from the coronal field restructuring \citep{hudson2000,fisher2012}.

How do BPs evolve in response to a major eruption? \citet{fan2007} performed magnetohydrodynamic (MHD) simulations for the eruption of a kink-unstable and a torus-unstable flux rope with BPs underneath. In the former case, the writhing motion induces internal reconnection that splits the rope vertically. As the top part escapes, the bottom part remains attached to the surface, and the BP survives. In the latter case, the expansion of the rope stretches the BP separatrix upward. Significant reconnection occurs at the BP separatrix until it finally disappears. Relevant observation, owing to the lack of high-cadence vector data, has been scarce.

NOAA AR 12673 (Figure~\ref{fig:bpevo}) produced the most intense flare of solar cycle 24 on September 6, 2017. The \textit{GOES} X9.3 flare was accompanied by a fast coronal mass ejection \citep{shencl2018} and white-light emission \citep{svanda2018}. Many studies have probed the coronal field structures prior to and during the eruption \citep{yangsh2017,zoup2020,inoue2021}. \citet{petrie2019} compared a pre-flare and a post-flare HMI vector magnetogram, and found clear signatures of magnetic imprints. \citet{jiang2018} performed data-constrained MHD modeling using HMI data, the results of which will be discussed in Section~\ref{sec:discussion}.

Here we report on the BP evolution in AR 12673 during this major eruption. Using HMI high-cadence vector data, we quantify the rapid changes in a temporally resolved manner. Below, we describe the data and methods in Section~\ref{sec:data}, present the results in Section~\ref{sec:result}, and discuss the implications and limitations in Section~\ref{sec:discussion}.


\begin{figure*}[t!]
\centering
\includegraphics[trim={0 0 0 1cm},clip,width=0.47\textwidth]{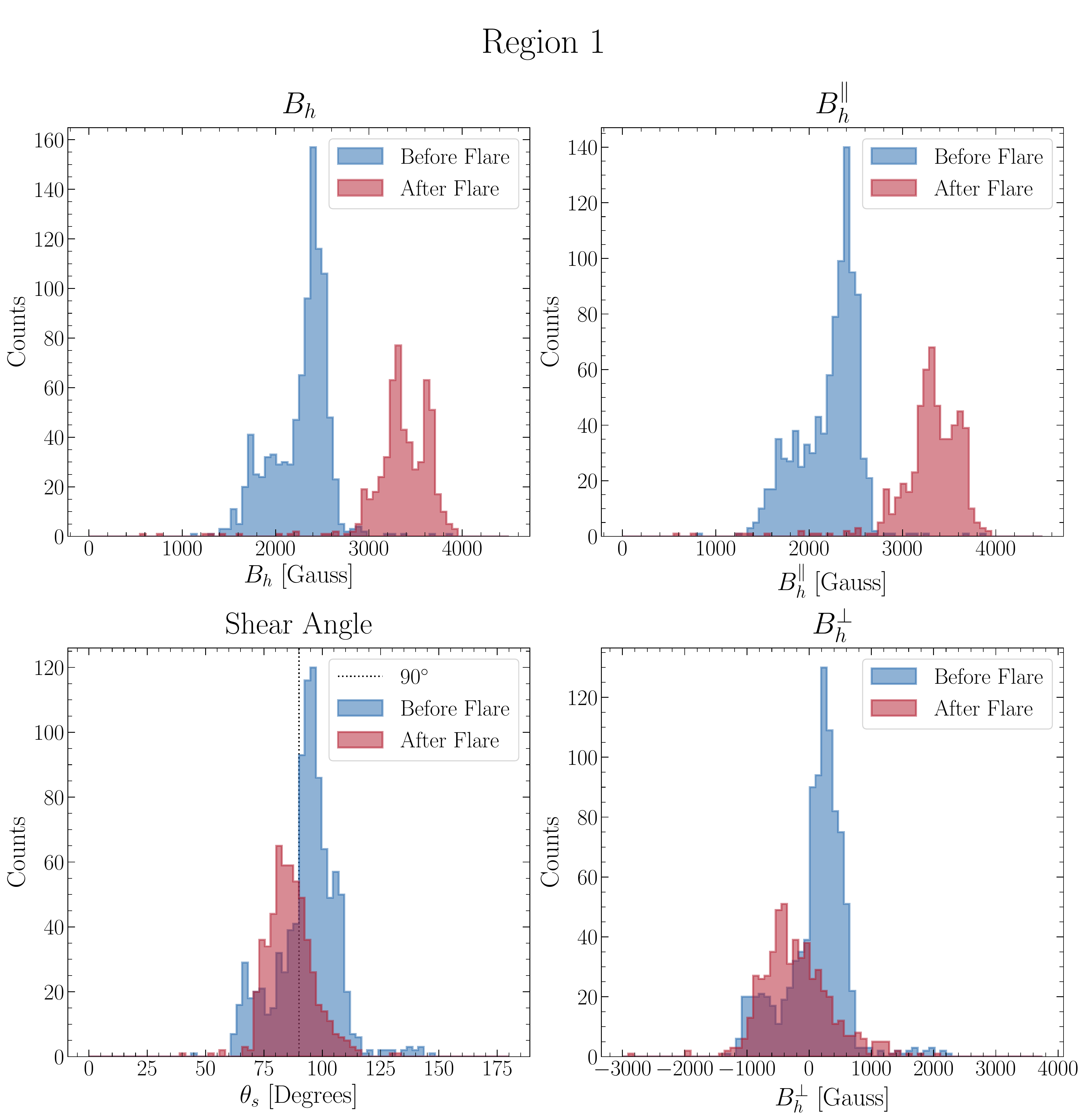}
\hspace{2mm}
\includegraphics[trim={0 0 0 1cm},clip,width=0.47\textwidth]{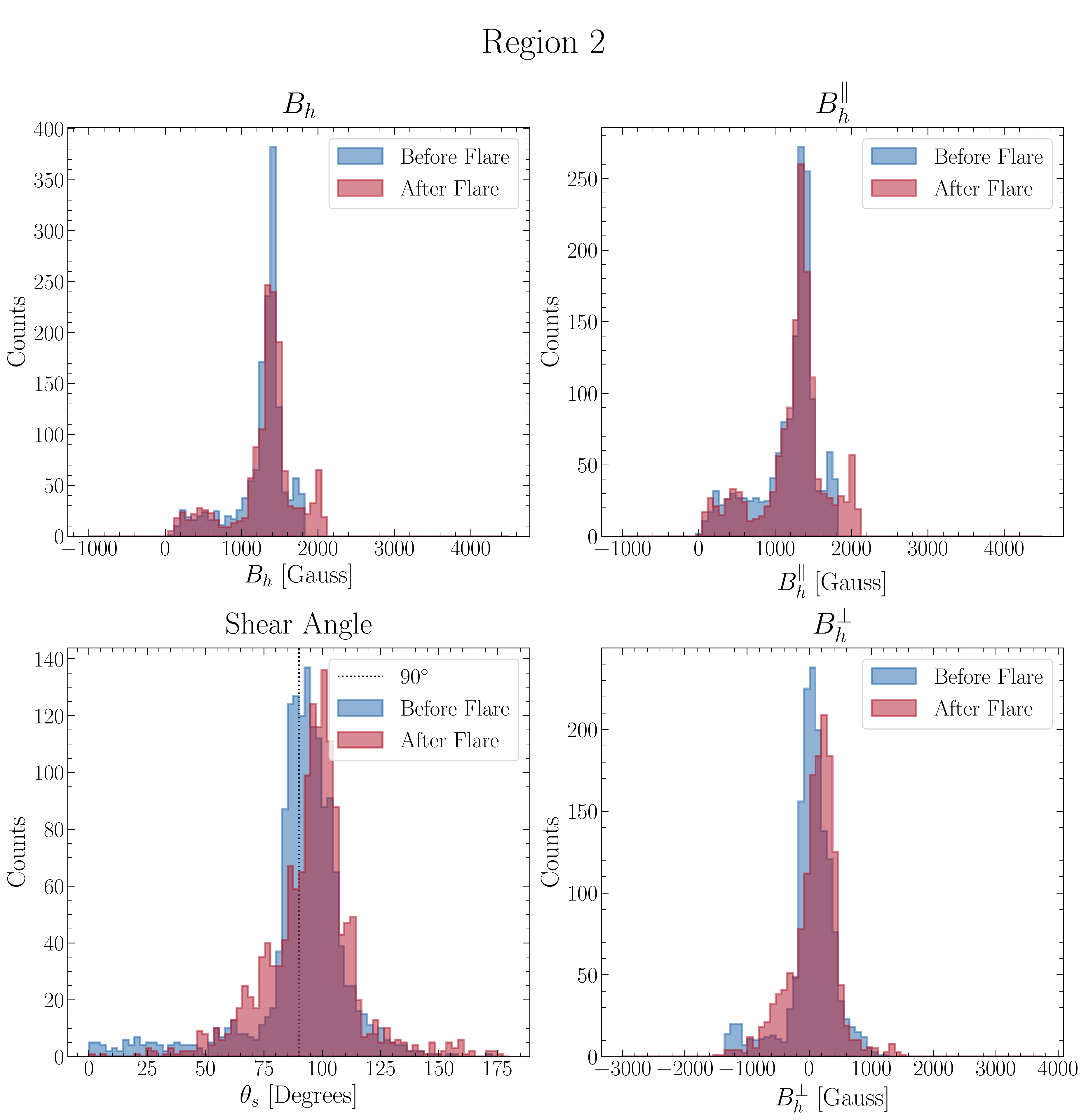}
\caption{Distributions of $B_h$, $B_h^\parallel$, $B_h^\perp$, and $\theta_s$ for 17 combined time steps before (blue, 11:27--11:51 UT) and after (red, 12:12--12:36 UT) the flare at the PIL pixels in Region 1 (left) and Region 2 (right). The vertical dotted line is for $\theta_s=90\degr$.}
\label{fig:Bhist}
\end{figure*}


\section{Data \& Methods}
\label{sec:data}

The \textit{GOES} X9.3 flare (\texttt{SOL2017-09-06T11:53}) has a start, peak, and end time in soft X-ray at 11:53 UT, 12:02 UT, and 12:10 UT, respectively. We analyze 88 frames of HMI vector magnetograms with a $90$~s cadence, which span roughly two hours centered around the flare peak. For each time step, we extract a $175$$\times$$175$ pixel map in a cylindrical equal area (CEA) coordinate \citep{sun2013}, centered at Carrington longitude $116\fdg9$ and latitude $-9\fdg3$ with a $0\fdg03$ sampling ($\sim$$360$~km). We adopt a local Cartesian approximation and decompose the field vectors as $\bm{B}=B_{x}\hat{\bm{x}} + B_{y} \hat{\bm{y}} + B_{z} \hat{\bm{z}}$, where $\hat{\bm{x}}$, $\hat{\bm{y}}$, and $\hat{\bm{z}}$ are the unit vectors pointing west, north, and upward, respectively.

We use the zero vertical field contour, $B_z=0$~G, to locate the ``PIL pixels''. For $B_z$ given at cell centers, the contouring routine typically returns coordinates on the cell edges. We identify the ``BP pixels'' on the PIL using the criterion \citep{titov1993}
\begin{equation}
\left.F\right | _{B_z=0} = \left. \bm{B}_h \cdot \bm{\nabla}_h B_z\, \right | _{B_z=0} > 0,	
\label{eqn:titov_car}
\end{equation}
where $\bm{B}_h=B_{x}\hat{\bm{x}} + B_{y} \hat{\bm{y}}$ is the horizontal field vector, and $\bm{\nabla}_h = {\partial}_x \hat{\bm{x}} + {\partial}_y \hat{\bm{y}}$ represents the horizontal gradient operator. We note that $\bm{\nabla}_{h} \, B_{z}$ is always perpendicular to the contours of $B_z$ and points in the direction of maximal increase in $B_z$. This means that $\left. \bm{\nabla}_{h} \, B_{z} \right|_{B_z=0}$ is perpendicular to the PIL, i.e., the $B_z=0$ contour, pointing from negative to positive polarity. Here and after, we assume the sub-cell values are bilinear, consistent with the contouring routine.

We calculate the following variables of interest for the PIL pixels: the horizontal field strength ($B_h$), the parallel component of $\bm{B}_h$ with respect to the PIL ($B_{h}^{\parallel}$), the corresponding, perpendicular component ($B_{h}^{\perp}$), and the magnetic shear angle ($\theta_s$). They are defined as follows:
\begin{equation}
B_{h} = \left( B_x^2+ B_y^2 \right)^{1/2},
\label{eqn:bh}
\end{equation}
\begin{equation}
B_{h}^{\perp} = \bm{B}_{h} \cdot \dfrac{\bm{\nabla}_{h} \: B_{z}}{\lVert \bm{\nabla}_{h} \: B_{z} \rVert},
\label{eqn:bh_perp}
\end{equation}
\begin{equation}
B_{h}^{\parallel} = \left[ B_{h}^{2} - \left( B_{h}^{\perp} \right) ^2 \right]^{1/2},
\label{eqn:bp_para}
\end{equation}
\begin{equation}
\theta_s  = 180\degr -  \mathrm{arccos} \: \dfrac{\bm{B}_{h} \cdot \bm{\nabla}_{h} \: B_{z}}{\lVert \bm{B}_{h} \rVert \lVert \bm{\nabla}_{h} \: B_{z} \rVert}.
\label{eqn:thetas}
\end{equation}
In our convention, $B_h^\perp$ is positive for BPs, and negative otherwise. The shear angle $\theta_s$ is zero if $\bm{B}$ is perpendicular to the PIL and points from the positive to the negative polarity. Sheared arcades have $0\degr<\theta_s<90\degr$; BPs have $\theta_s > 90\degr$. We do not consider $B_z$ as all analysis are restricted to the PIL pixels, where $B_z=0$~G is satisfied to machine accuracy.


\begin{figure*}[t!]
\centering
\includegraphics[trim={0 0 0 1cm},clip,width=0.47\textwidth]{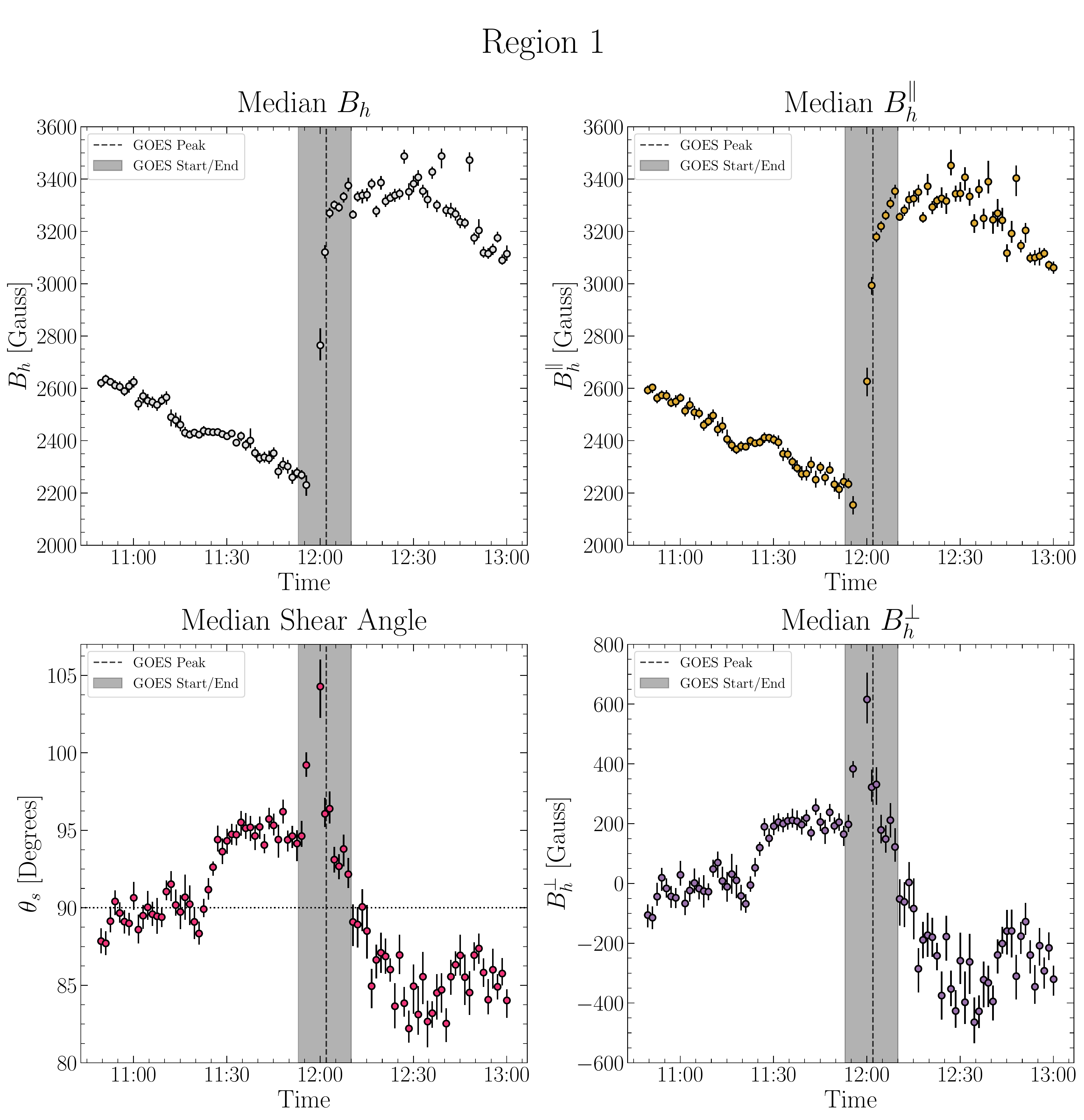}
\hspace{2mm}
\includegraphics[trim={0 0 0 1cm},clip,width=0.47\textwidth]{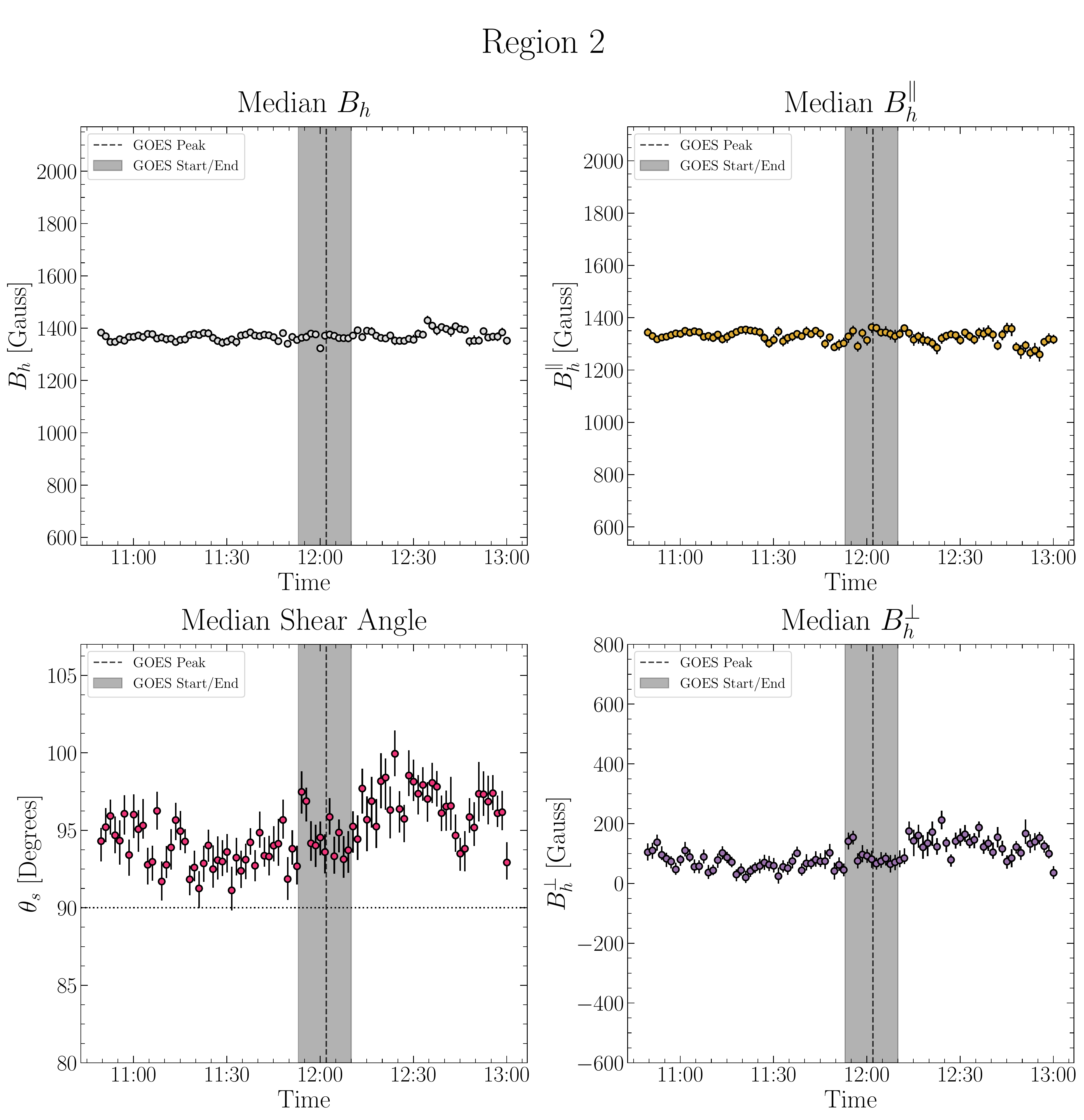}
\caption{Time series of median $B_h$, $B_h^\parallel$, $B_h^\perp$, and $\theta_s$ at PIL pixels in Region 1 (left) and Region 2 (right). The error bars show the 1$\sigma$ confidence interval of the median derived from the Monte-Carlo method. The vertical gray bar (dotted line) shows the flare duration (peak time). The horizontal dotted line is for $\theta_s=90\degr$.}
\label{fig:Bts}
\end{figure*}


Following \citet{avallone2020}, we use a Monte-Carlo method to estimate the statistical uncertainties. A random sample ($N=100$) is drawn for $\bm{B}$ at each pixel in the full-disk magnetogram based on the variances and covariances provided by the HMI pipeline. It is then used to calculate a sample of the desired variables. The $1\sigma$ confidence interval is finally quoted.

For analysis, we exclude those pixels with weaker field strength, $\lVert\bm{B}\rVert<200$~G, as they have relatively higher noises. Several sub-regions in AR 12673 exhibit particularly large uncertainties, including the sunspot umbra in the north, the $\delta$-penumbra at the center, and the flare ribbons. We identify and exclude them by imposing an empirical threshold (two standard deviations above the median) on the statistical uncertainties of the three field components: $\sigma_{B_x}>162$~G, $\sigma_{B_y}>152$~G, $\sigma_{B_z}>135$~G. We further discuss the uncertainties in Section~\ref{sec:discussion}.

Using the flare ribbon information from the \verb+RibbonDB+ database \citep{kazachenko2017} derived from the the Atmospheric Imaging Assembly \citep{lemen2012} 1600~{\AA} images, we locate all the pixels that have risen above 8 times the median image intensity up to the time of interest. We consider them as the foot points of the field lines that have reconnected, and integrate the corresponding HMI $B_z$ pixels to estimate the accumulated reconnecting magnetic flux $\Phi$:
\begin{equation}
\Phi  = \int \left| B_z \right| \mathrm{d}\,S_\text{ribbon},
\label{eqn:phi}
\end{equation}
where $S_\text{ribbon}$ denotes the accumulated ribbon area. Equation~(\ref{eqn:phi}) includes both positive and negative fluxes. Different thresholds for the ribbon brightness, i.e. 6 and 10 times the median, are used for the uncertainty estimate.


\begin{figure*}[t!]
\centering
\includegraphics[width=0.44\textwidth]{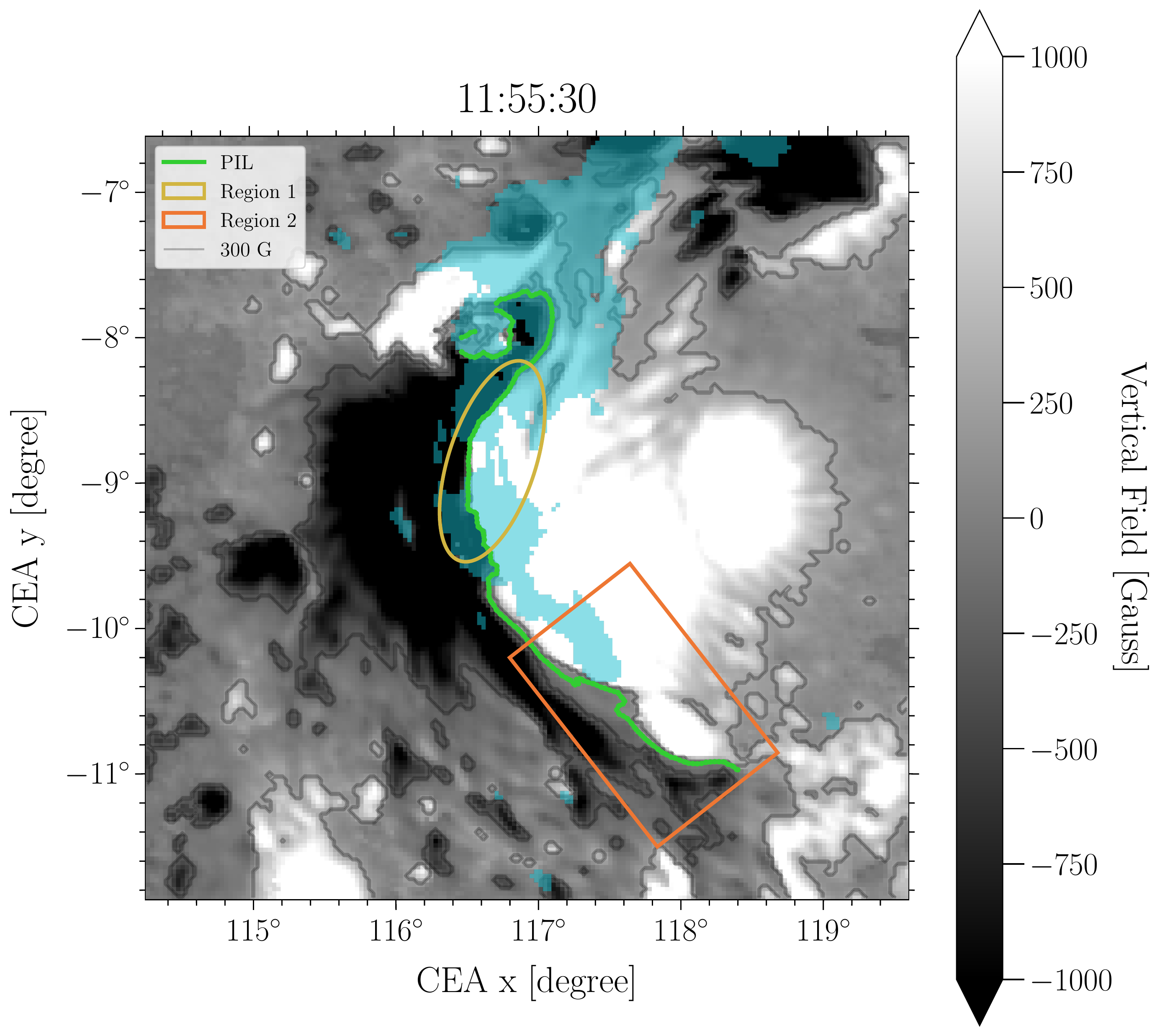}
\hspace{8mm}
\includegraphics[width=0.44\textwidth]{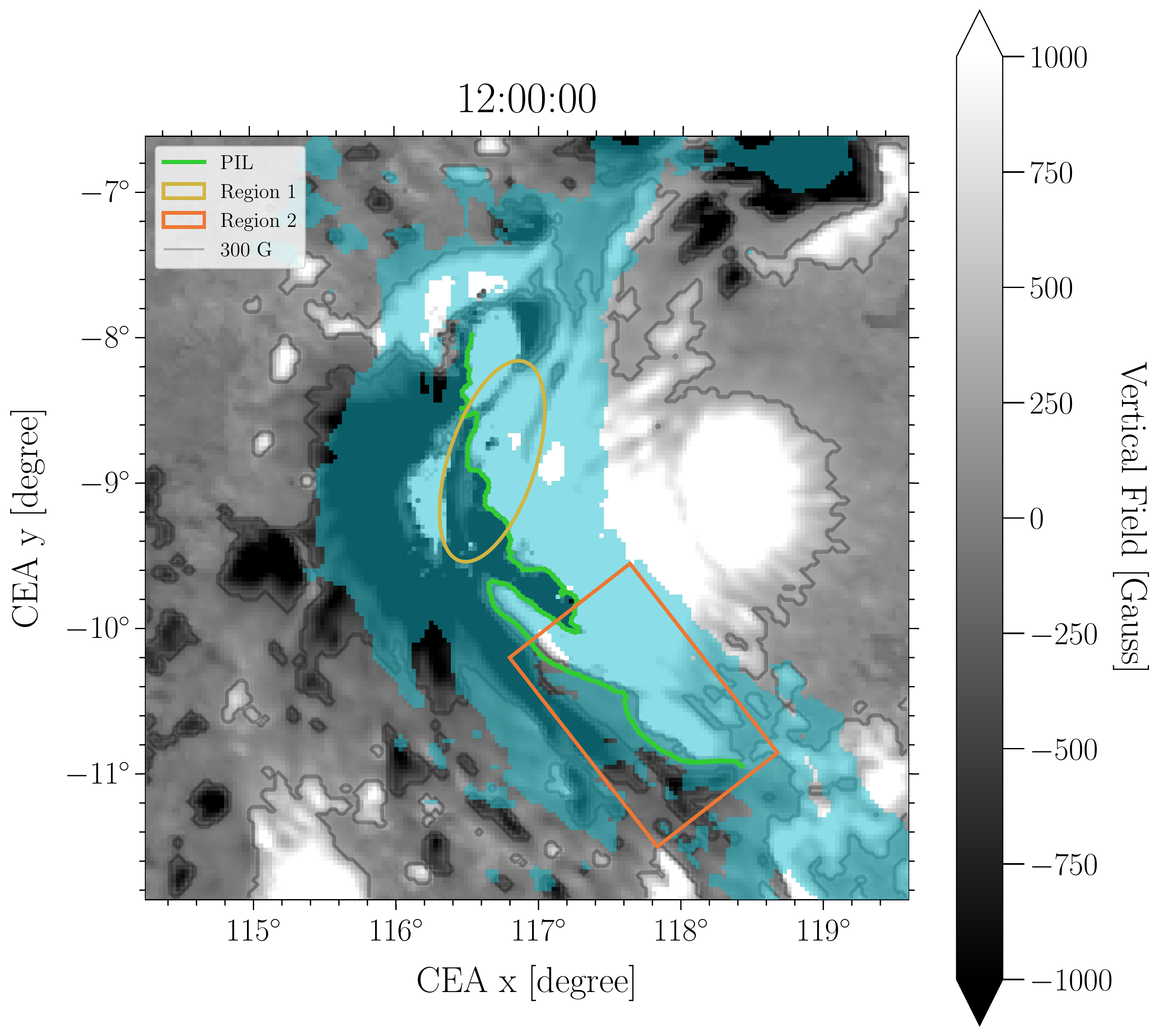}
\caption{AIA 1600~{\AA} flare ribbons in teal during the flare onset (left, 11:55:27 UT) and shortly before the flare peak (right, 12:00:15 UT). (An animation of this figure is available via \href{https://doi.org/10.5281/zenodo.5585623}{Zenodo}.)}
\label{fig:flareRB}
\end{figure*}


\begin{figure*}[t!]
\centering
\includegraphics[width=0.75\textwidth]{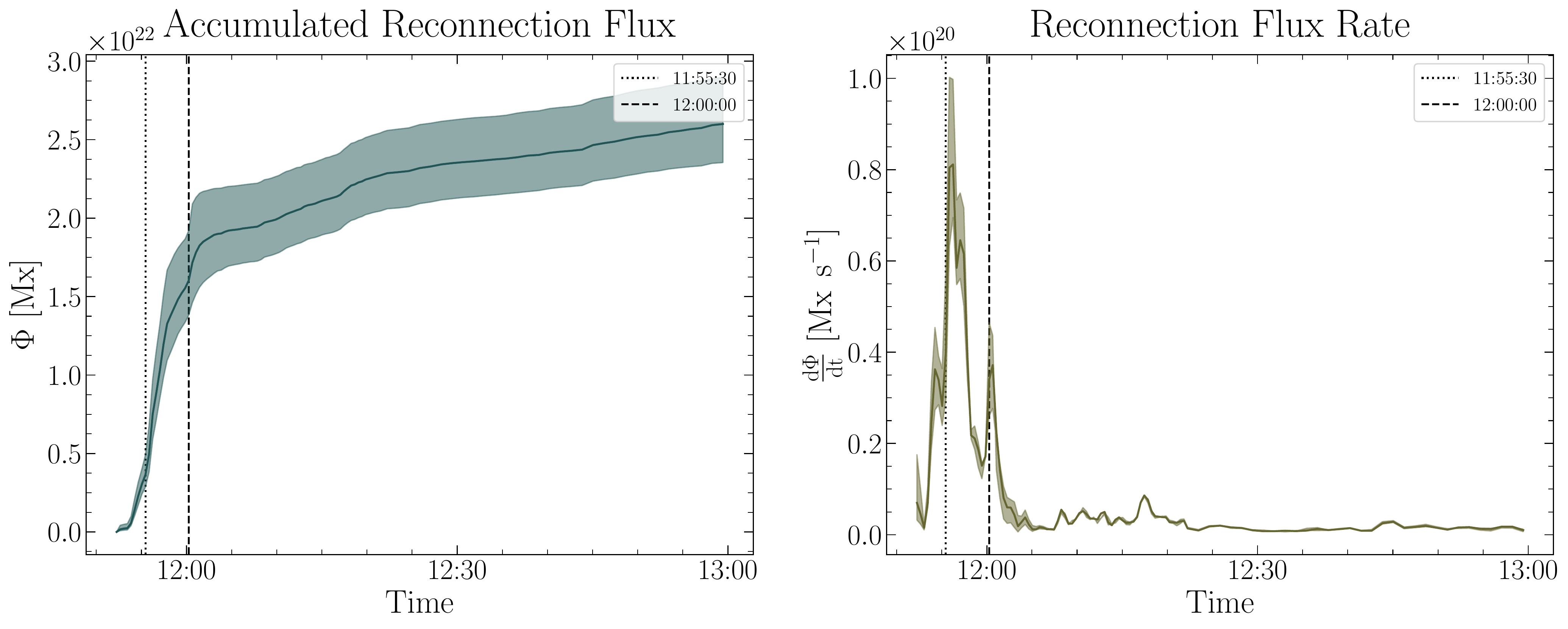}
\caption{Left: accumulated reconnection flux $\Phi$. Right: reconnection flux rate $\mathrm{d}\Phi/\mathrm{d}t$. The vertical dotted and dashed lines indicate the two times shown in Figure~\ref{fig:flareRB}.}
\label{fig:recflux}
\end{figure*}


\section{Result}
\label{sec:result}

AR 12673 hosted a complex $\beta\gamma\delta$-class sunspot group prior to the eruption. The coronal loops in the core region have an inverse-S shape. This suggests that the coronal fields have left-handed twist, consistent with the result from magnetic extrapolation \citep[e.g.,][]{wangr2018}.

For each time step, we calculate $F$ with Equation (\ref{eqn:titov_car}) and other variables of interest for all the PIL pixels with low statistical uncertainty. We select two subregions, Region 1 and 2  (Figure~\ref{fig:bpevo}), for further analysis. We exclude the data taken during the flare (12:12--12:36 UT), because the flare emission creates significant transient signal at ribbon loci (see Section~\ref{sec:discussion}).

For Region 1 at the central portion of the PIL, $73\%$ of pixels ($43$ out of $59$) have $F>0$ at 11:37:30 UT (pre-flare) and are identified as BP. This is in sharp contrast with the 12:25:30 UT (post-flare) frame, where only $39\%$ of pixels ($14$ out of $36$) are identified as BP. As directly discernible in Figure~\ref{fig:bpevo}, the BP appears to have ``disintegrated'' during the eruption. For Region 2 at the southern portion of the PIL, $63\%$ of pixels ($55$ out of $87$) are identified as BPs pre-flare, and $71\%$ of pixels ($57$ out of $80$) post-flare. The BP appears to have survived the eruption.

The distributions of several variables of interest are shown in Figure~\ref{fig:Bhist}. In Region 1, both $B_{h}$ and $B_{h}^{\parallel}$ greatly increased after the flare; the median increase is $988$~G for $B_{h}$ and $1024$~G for $B_{h}^{\parallel}$. The significant increase is in line with previous findings of the magnetic imprints. In contrast, $B_{h}^{\perp}$ reversed sign; its median changed from $201$~G to $-262$~G. The median $\theta_s$ decreased by $9\degr$, from $95\degr$ to $86\degr$. In Region 2, most pixels had $B_{h}^{\perp}>0$~G and $\theta_s>90\degr$ both before and after the flare. The median change before to after the flare are $9$~G, $3$~G, $79$~G, and $4\degr$ for $B_h$, $B_h^\parallel$, $B_h^\perp$, and $\theta_s$, respectively.

We conduct a two-sample Kolmogorov-Smirnov (K-S) test to evaluate whether the pre- and post-flare variables are likely drawn from the same distribution. The K-S $D$ statistics for Region 2 are $3$--$10$ times smaller than those of Region 1. Nevertheless, the $p$ values for all four variables are small ($p<10^{-6}$) in both subregions, which indicates some differences between the pre- and post-flare states.

The temporal profiles of the median values are shown in Figure~\ref{fig:Bts}. As expected, Region 1 exhibited a rapid and permanent increase of roughly 1~kG in $B_{h}$ during the flare, which mostly comes from $B_{h}^\parallel$. Interestingly, the median $B_{h}^\perp$ ($\theta_s$) increased from about $0$~G ($90\degr$) to $200$~G ($95\degr$) between 11:20 UT and 11:35 UT, suggesting rapid formation of new BPs. Their values subsequently decreased after the flare onset, which caused the apparent BP disintegration. In comparison, the temporal variations in Region 2 were much smaller.

We show the flare ribbon evolution in Figure~\ref{fig:flareRB} and the estimated reconnection flux $\Phi$ in Figure~\ref{fig:recflux}. At the flare onset, the ribbons first appeared close to the northern portion of the PIL, including Region 1 (left panel of Figure~\ref{fig:flareRB}). This suggests that the BP separatrix in Region 1 may have participated in the flare reconnection early on, when the reconnection flux rate $\mathrm{d}\Phi/\mathrm{d}t$ was near its maximum, $8.11\times10^{19}$~Mx~s$^{-1}$. The ribbons subsequently extended along the PIL to cover Region 2 (right panel), when $\mathrm{d}\Phi/\mathrm{d}t$ was much reduced, $3.72\times10^{19}$~Mx~s$^{-1}$. In the meanwhile, the ribbons also moved away from the PIL, suggesting that the coronal reconnection proceeded to higher altitudes. The evolution pattern is typical for two-ribbon flares \citep{qiu2009}.


\section{Discussion}
\label{sec:discussion}

Our observations show that a segment of BP in AR 12673 rapidly disintegrated during the X9.3 flare. The apparent disintegration is a consequence of the changing $B_h^\perp$ and $\theta_s$, as the local magnetic azimuth angles rotated by about $10\degr$ over half an hour. The post-flare $\bm{B}_h$ pointed from the positive to the negative polarity, and the low-lying field lines assumed a sheared arcade topology instead. The rotation is reminiscent of a $12\degr$ permanent rotation of a sunspot \citep{liuc2016} and a $12$--$20\degr$ transient rotation of photospheric fields \citep{xuy2018} during flares.

Our observations also show that the parallel component of the horizontal field increased by about $1$~kG, in line with previous reports of magnetic imprint. This is consistent with the findings of \citet{petrie2019} on Region 1.

The morphological evolution of the flare ribbons contains rich information about the coronal reconnection processes. During the early phase of the flare, the ribbons extended along and moved away from the PIL. The extension is approximately centered at Region 1; it is mainly northward (southward) in the negative (positive) polarity, manifesting the sheared component of the reconnecting fields \citep{qiu2009}. The outward motion away from the PIL indicates that the reconnection occurs at progressively higher altitude, and starts to involve the overlying fields. We note that the ribbon evolution after the flare peak was rather complex, with possible inward motions toward the PIL near Region 1. Given the multi-polar nature of AR 12673 and the possible multi-flux-rope eruption \citep{hou2018}, the peculiar motion may be a consequence of breakout-type reconnection \citep[e.g.,][]{lynch2013,dahlin2019}. A detailed analysis is deferred to future work.

In \citet{fan2007}, the BP separatrix disappears due to intense reconnection that proceeds to very low altitudes. Our observations appear to be consistent with the scenario, especially when the contrasting behaviors of Region 1 and 2 are taken into account. The early, more intense reconnection likely occurred at a lower altitude above Region 1. It may have involved the entire BP separatrix, leading to the rapid disintegration of the BP there. Region 2 was involved later when the reconnection proceeded to higher altitudes, as suggested by the separation of the ribbons away from the PIL. The BP separatrix there was less involved, and the BP survived. Indeed, photospheric white-light flare ribbons were observed by HMI, and the $10$--$25$~keV hard X-ray kernels were observed by the RHESSI satellite \citep{yangsh2017,romano2019}. Both appeared near Region 1 only and are indicative of intense magnetic reconnection.

\citet{jiang2018} extrapolate a nonlinear force-free field from HMI vector data as the initial state of their MHD simulation. The pre-eruption coronal field contains a low-lying flux rope right above the BP in Region 1. The small residual Lorentz force in the extrapolation model provides the initial disturbance; plasma instability eventually leads to a full eruption. Similar to \citet{fan2007}, an intense current sheet forms at the stretched BP separatrix, which is eventually converted into sheared arcades. The ribbon motions during the early stage of the flare, i.e., extension along and expansion away from the PIL, are well reproduced in the model. Even though the photospheric $\bm{B}_h$ is not observationally constrained in the numerical simulation of \citet{jiang2018} for AR 12673, the simulated BP evolution appears to match the HMI data qualitatively, including a clear increase in the strength of the horizontal field, $B_h$ (C. Jiang, private communication).

We note that many previous observational and modeling studies (Section~\ref{sec:intro}) have discussed BP separatrix reconnection, though mostly in less energetic contexts such as H$\alpha$ surges or confined flares. The photospheric field changes in those events remain to be studied, but are likely less drastic.

We finally comment on the uncertainties. First, the $\delta$-penumbra at the southern section of Region 1 (centered at $(116\fdg5,-9\fdg5)$ in Figure~\ref{fig:bpevo}) consistently shows high statistical errors, therefore, reducing the number of PIL pixels considered for analysis. The cause may be the combined effect of extremes in the Doppler velocity and limitations of the magnetic inference technique. Second, the average $\sigma_{B_x}$, $\sigma_{B_y}$, and $\sigma_{B_z}$ in pixels where $\lVert\bm{B}\rVert>200$~G increased by $36$~G, $26$~G, and $30$~G respectively post-flare, leading to a reduced number of pixels available for analysis. The increase mainly originates from the northern umbra and the $\delta$-penumbra, the reason for which is unclear. The regions swept by the flare ribbons, on the other hand, have similar uncertainty as the surrounding regions in the post-flare frames. Third, flare emission in ribbons are known to cause transient artifact in photospheric magnetogram such as sign reversals \citep{qiu2003,sun2017a}. The effect clearly manifests in the right panel of Figure~\ref{fig:flareRB} as parasitic polarities that distort the PIL. They are also visible in the animation of Figures~\ref{fig:bpevo} and~\ref{fig:flareRB}. Fourth, the determination of the BP relies on the resolution of the $180$-degree ambiguity of the magnetic azimuth. The HMI pipeline algorithm \citep{leka2009} includes the divergence-free constraint, which has been shown to work well in several theoretical cases with BPs \citep{lij2007}. Additional constraints come from the inverse-S-shaped sigmoids observed in the corona, which is consistent with the negative helicity inferred from the photosphere vector field. Finally, flare ribbon observations in $1600$~{\AA} are subject to saturation during larger flares. The associated ``blooming'' effect, where the saturation causes excess charge to spread to neighboring pixels on the detector, can lead to an overestimate of the ribbon area and the absolute values of the reconnection flux rates. Diffraction patterns from the AIA entrance filter may also affect the result.
 

\begin{acknowledgments}
We thank Chaowei Jiang for helpful discussions. J.L. and X.S. are supported by NSF awards \#1848250, \#1854760, and the state of Hawai`i. M.D.K. acknowledges support from NASA ECIP award NNH18ZDA001N. The \textit{SDO} data are courtesy of NASA, the HMI, and the AIA science teams.
\end{acknowledgments}

\facilities{\textit{SDO}}



\end{CJK*}


\bibliographystyle{apj_url}
\bibliography{bpevo}

\end{document}